\documentclass[a4paper,12pt]{article}
\usepackage[pctex32]{graphics}

\begin{document}
\topmargin 0pt \oddsidemargin 0mm

\renewcommand{\thefootnote}{\fnsymbol{footnote}}
\begin{titlepage}
\begin{flushright}
INJE-TP-04-08\\
hep-th/0412224
\end{flushright}

\vspace{5mm}
\begin{center}
{\Large \bf Holographic principle and dark energy} \vspace{12mm}

{\large  Yun Soo Myung\footnote{e-mail
 address: ysmyung@physics.inje.ac.kr}}
 \\
\vspace{10mm} {\em  Relativity Research Center and School of
Computer Aided Science, Inje University, Gimhae 621-749, Korea}
\end{center}

\vspace{5mm} \centerline{{\bf{Abstract}}}
 \vspace{5mm}
We discuss  the relationship between  holographic entropy bounds
and  gravitating systems. In order to obtain a holographic energy
density,  we introduce the Bekenstein-Hawking entropy $S_{\rm BH}$
and its corresponding energy $E_{\rm BH}$ using the Friedman
equation. We show that the holographic energy bound proposed by
Cohen {\it et al} comes from the Bekenstein-Hawking bound for a
weakly gravitating system. Also we find that  the holographic
energy density with the future event horizon deriving an
accelerating universe  could be given by vacuum fluctuations of
the energy density.
\end{titlepage}

\newpage
\renewcommand{\thefootnote}{\arabic{footnote}}
\setcounter{footnote}{0} \setcounter{page}{2}
\section{Introduction}
Supernova (SN Ia) observations suggest that our universe is
accelerating and the dark energy contributes $\Omega_{\rm
DE}\simeq 0.60-0.70$ to the critical density of the present
universe~\cite{SN}. Also  cosmic microwave background (CMB)
observations~\cite{Wmap} imply that the standard cosmology is
given by the inflation and FRW universe~\cite{Inf}. A typical
candidate for the dark energy is the cosmological constant.
Recently Cohen {\it et al} showed that in quantum field theory, a
short distance cutoff (UV cutoff: $\Lambda$) is related to a long
distance cutoff (IR cutoff: $L_{\rm \Lambda}$) due to the limit
set by forming a black hole~\cite{CKN}. In other words, if
$\rho_{\rm \Lambda}$ is the quantum zero-point energy density
caused by a UV cutoff $\Lambda$, the total energy of the system
with size $L_{\rm \Lambda}$ should not exceed the mass of the same
size-black hole: $L_{\rm \Lambda}^3 \rho_{\rm \Lambda}\le L_{\rm
\Lambda} M_p^2$ with the Planck mass of $M_p^2=1/G$. The largest
$L_{\rm \Lambda}$ is chosen as the one saturating this inequality
and its holographic energy density is then given by $\rho_{\rm
\Lambda}= 3c^2M_p^2/8\pi L_{\rm \Lambda}^2$ with a numerical
factor $3c^2$.  Taking $L_{\rm \Lambda}$ as the size of the
present universe, the resulting energy  is comparable to the
present dark energy~\cite{HMT}. Even though this holographic
approach leads to the data, this description is incomplete because
it fails to explain the dark energy-dominated present
universe~\cite{HUE}. In order to resolve this situation, one is
forced to introduce another candidates for  IR cutoff. One  is the
particle horizon $R_{\rm H}$ which was used in the holographic
description of cosmology by Fischler and Susskind~\cite{FS}. This
gives $\rho_{\rm \Lambda} \sim a^{-2(1+1/c)}$ which implies
$\omega_{\rm H}>-1/3$~\cite{LI}. This corresponds to a
decelerating universe and unfortunately  is not our case. In order
to find an accelerating universe, we need the future event horizon
$R_{\rm h}$. With $L_{\rm \Lambda}=R_{\rm h}$ one finds $\rho_{\rm
\Lambda} \sim a^{-2(1-1/c)}$ to describe the dark energy with
$\omega_{\rm h}<-1/3$. This is close enough to $-1$ to agree with
the data~\cite{SN}. However, this relation seems to be rather ad
hoc chosen and one has to justify whether or not $\rho_{\rm
\Lambda}= 3c^2M_p^2/8\pi L_{\rm \Lambda}^2$ is appropriate to
describe the present universe.

 On the other hand, the implications of the
cosmic holographic principle  have been investigated in the
literature~\cite{Hooft,Beke,FS,Hubb,Bous}. However, these focused
on the decelerating universe, especially for a radiation-dominated
universe.

 In this letter we will clarify  how
the cosmic holographic principle could be used for obtaining the
holographic energy density. This together with the future event
horizon is a candidate for the dark energy to derive an
accelerating universe. Further we wish to seek the origin of the
holographic energy density.

\section{Cosmic holographic bounds}
We briefly review the cosmic  holographic bounds for our  purpose.
Let us start an $(n+1)$-dimensional Friedman-Robertson-Walker
(FRW) metric
\begin{equation}
\label{2eq1} ds^2 =-dt^2 +a(t)^2 \Big[ \frac{dr^2}{1-kr^2} +r^2
d\Omega^2_{n-1} \Big],
\end{equation}
where $a$ is the  scale factor of the universe and
$d\Omega^2_{n-1}$ denotes the line element of an
$(n-1)$-dimensional unit sphere. Here $k=-1,~0,~1$ represent that
the universe  is open, flat, closed, respectively. A cosmological
evolution is determined by the two Friedman equations
\begin{eqnarray}
\label{2eq2}
 && H^2 =\frac{16\pi G_{n+1}}{n(n-1)}\frac{E}{V}
-\frac{k}{a^2}, \nonumber \\
&& \dot H =-\frac{8\pi G_{n+1}}{n-1}\left (\frac{E}{V} +p\right)
    +\frac{k}{a^2},
\end{eqnarray}
where $H$ represents the Hubble parameter with the definition
$H={\dot a}/a$ and the overdot stands for  derivative with respect
to the cosmic time $t$,  $E$ is the total energy of matter filling
the universe, and $p$ is its pressure. $V$ is the volume of the
universe, $V=a^n \Sigma^n_k$ with $\Sigma^n_k$ being the volume of
an $n$-dimensional space with a curvature constant $k$, and
$G_{n+1}$ is the Newton constant in ($n+1$) dimensions. Here we
assume the equation of state:
 $p=\omega \rho,~ \rho=E/V$.
First of all, we introduce  two  entropies
 for the holographic description of a universe~\cite{Verl,SV}:
\begin{equation}
\label{2eq3}
  S_{\rm
 BV}=\frac{2\pi}{n}Ea,
  ~~ S_{\rm
 BH}=(n-1)\frac{V}{4G_{n+1}a},
\end{equation}
where $S_{\rm
 BV}$ and $S_{\rm
 BH}$ are called the Bekenstein-Verlinde entropy and Bekenstein-Hawking entropy, respectively.
Then, the first Friedman equation can be rewritten as
\begin{equation}
\label{Fried} (Ha)^2=2\frac{S_{\rm BV}}{S_{\rm BH}}-k.
\end{equation}
 We  define a quantity $E_{\rm BH}$ which
corresponds to energy needed to form a universe-sized black hole
by analogy with $S_{\rm BV}$: $ S_{\rm BH}=(n-1)V/4G_{n+1}a \equiv
2\pi E_{\rm BH} a/n $. Using these, for $Ha \le \sqrt{2-k}$, one
finds the Bekenstein-Hawking bound for a weakly self-gravitating
system as

\begin{equation}
\label{WB} E\le E_{\rm BH} \leftrightarrow S_{\rm BV} \le S_{\rm
BH},
\end{equation}
while for $Ha \ge \sqrt{2-k}$, one finds the cosmic holographic
bound for a strongly self-gravitating system as

\begin{equation}
\label{SB} E\ge E_{\rm BH} \leftrightarrow S_{\rm BV} \ge S_{\rm
BH}.
\end{equation}

\section{Holographic energy bounds}
 First we study how the gravitational holography goes well with a (3+1)-dimensional effective
theory. For convenience we choose  the volume of the system as
$V_{\rm \Lambda}=4\pi L_{\rm \Lambda}^3/3 \sim L_{\rm \Lambda}^3$.
For an effective quantum field theory in a box of volume $V_{\rm
\Lambda}$ with a UV cutoff $\Lambda$\footnote{Precisely, $M_{\rm
\Lambda}$ is  more suitable for an UV cutoff than $\Lambda$, but
we here use the latter instead of $M_{\rm \Lambda}$ for
convenience.}, its entropy scales extensively as~\cite{CKN}
 \begin{equation}
 \label{2eq8}
 S_{\rm \Lambda} \sim L_{\rm \Lambda}^3\Lambda^3.
 \end{equation}
 However, the Bekenstein postulated that the maximum
entropy in a box of volume $V_{\rm \Lambda}$ behaves
non-extensively, growing only as the enclosed area $A_{\rm
\Lambda}$ of the box. We call it  the gravitational holography.
The Bekenstein entropy bound is satisfied in the effective theory
if
\begin{equation}
\label{BB} S_{\rm \Lambda} \sim L_{\rm \Lambda}^3\Lambda^3 \le
S_{\rm BH} \equiv\frac{2}{3}\pi M^2_p L_{\rm \Lambda}^2  \sim
M^2_p L_{\rm \Lambda}^2,
\end{equation}
where $S_{\rm BH}$ is the closest one which comes to the usual
form of Bekenstein-Hawking entropy $A_{\rm \Lambda}/4G=\pi M^2_p
L_{\rm \Lambda}^2 $ for a black hole of radius $L_{\rm \Lambda}$.
Thus the IR cutoff cannot be chosen independently of the UV cutoff
$\Lambda$, if we introduce the gravitational holography. It scales
like  $L_{\rm \Lambda} \sim \Lambda^{-3}$. This bound is suitable
for the system with a relatively low energy density. However, an
effective theory that can saturate the inequality of Eq.(\ref{BB})
includes many states with the Schwarzschild radius $L_{\rm
S}=2GM_{\rm S}$ much larger than the size of a box $L_{\rm
\Lambda}$ $(L_{\rm S}> L_{\rm \Lambda})$.  To avoid this
difficulty, one proposes a rather strong constraint on the IR
cutoff which excludes all states that lie within the Schwarzschild
radius. Then one finds  cases with  $L_{\rm S}< L_{\rm \Lambda}$.
Since the maximum energy density $\rho_{\rm \Lambda}$ in the
effective theory is $\Lambda^4$, the total energy scales as $
E_{\rm \Lambda}=V \rho_{\rm \Lambda} \sim L_{\rm \Lambda}^3
\Lambda^4$. As a result, the constraint on the IR cutoff is given
by
\begin{equation}
\label{EB} E_{\rm \Lambda} \sim L_{\rm \Lambda}^3\Lambda^4 \le
M_{\rm S} \sim L_{\rm \Lambda} M_p^2,
\end{equation}
where the IR cutoff scales as $L_{\rm \Lambda} \sim \Lambda^{-2}$.
This bound is more restrictive  than the Bekenstein bound in
Eq.(\ref{BB}). By definition, the two scales $\Lambda$ and $L_{\rm
\Lambda}$ are independent to each other initially. To reconcile
the breakdown of the quantum field theory to describe a black
hole, one proposes a relationship between UV and IR cutoffs.  An
effective field theory could  then describe a system including
even a black hole.

Now we explain the bound of Eq.(\ref{EB}) within our framework. We
wish to interpret it in view of cosmic holographic bounds. Here
$k=0$ and a physical scale  $ a \sim L$.  From Eq.(\ref{WB}), for
$L \le \sqrt{2}/H$, one finds the holographic bound for a weakly
self-gravitating system as $E \sim L^3\Lambda^4 \le E_{\rm
BH}\equiv 2LM_p^2 \sim LM^2_p$. Also this inequality is derived
from the Bekenstein-Hawking entropy bound: $S_{\rm BV} \le S_{\rm
BH}$. Here $S_{\rm BV}$ is not really as an entropy but rather as
the energy measured with respect to an appropriately chosen
conformal time coordinate~\cite{Verl}. Also the role of $S_{\rm
BH}$ is not to serve a bound on the total entropy, but rather on a
sub-extensive component of the entropy that is associated with the
Carsimir energy  $E_{\rm c}$ of the CFT. Consequently, the bound
of Eq.(\ref{EB}) comes from the Bekenstein-Hawking bound for a
weakly gravitating system. Furthermore, if $L=\sqrt{2}/H$, one
finds the saturation which states that $S_{\rm BV}=S_{\rm BH}
\leftrightarrow E=E_{\rm BH}$. We remind the reader that $E_{\rm
BH}$ is an energy to form a universe-sized black hole. The
universe is in a weakly self-gravitating phase when its total
energy $E$ is less than $E_{\rm BH}$, and in a strongly
gravitating phase for $E>E_{\rm BH}$. We emphasize that comparing
with the Bekenstein bound in Eq.(\ref{BB}), the Bekenstein-Hawking
bound in Eq.(\ref{EB}) comes out only when taking the Friedman
equation (dynamics) into account~\cite{SV}. Hence the
Bekenstein-Hawking bound is more suitable for cosmology than the
Bekenstein bound. Up to now, we consider the cosmic  holographic
bounds for the decelerating universe which includes either a
weakly gravitating system or a strongly gravitating one.

\section{Holographic dark energy}

In order to describe the dark energy, we have to choose a
candidate. There are many candidates. In this work we choose the
holographic energy to describe the dark energy. We take the
largest $L_{\rm \Lambda}$ as the one saturating the inequality of
Eq.(\ref{EB}). Then  we find a relation for the cosmological
energy density (cosmological constant): $\rho_{\rm
\Lambda}=3c^2M^2_p/8\pi L_{\rm \Lambda}^2$ with a numerical
constant $3c^2$.  In the case of $c=1$, it corresponds to a
variant of the cosmological constant because the conventional form
is usually given by $\tilde{\rho}_{\rm \Lambda} \sim
\tilde{\Lambda}^4 =1/\tilde{L}^{4}$ with $\tilde{\Lambda} \sim
1/\tilde{L}$ in the de Sitter thermodynamics\cite{GH,NOB,Myung}.

 Here  three choices are possible
 for $L_{\rm \Lambda}$~\cite{LI}. If one chooses IR cutoff
as the size of our universe ($L_{\rm \Lambda}=d_{\rm H}=1/H$), the
resulting energy is comparable to the present dark
energy~\cite{HMT}. Even though this holographic approach leads to
the data, this description is incomplete because it fails to
explain the present universe of dark-energy dominated phase with $
\omega=p/\rho \le -0.78$~\cite{HUE,HOV}. In this case the Friedman
equation including  a matter  of $\rho_m$ is  given by
$\rho_m=3(1-c^2)M_p^2H^2/8\pi$, which leads to the dark energy
with $\omega=0$. However, the accelerating universe requires
$\omega<-1/3$ and thus this case is excluded.
 In order to
resolve this situation, one is forced to introduce the particle
horizon $L_{\rm \Lambda}=R_{\rm H}=a \int_0^t (dt/a)=a
\int^a_0(da/Ha^2)$ which was used in the holographic description
of cosmology by Fischler and Susskind~\cite{FS}. In this case, the
Friedman equation of $H^2=8\pi \rho_{\rm \Lambda}/3M_p^2$ leads to
an integral equation  $HR_{H}=c$. Finally it takes the form of a
differential equation
\begin{equation}
\label{DEP}c\frac{d}{da}\Big(\frac{H^{-1}}{a}\Big)=\frac{1}{Ha^2}.
\end{equation}
It gives $\rho_{\rm \Lambda}\sim a^{-2(1+1/c)}$, which implies
$\omega_{\rm H}= -1/3(1-2/c)>-1/3$. This is still  a decelerating
phase because the comoving Hubble scale of  $H^{-1}/a$ is
increasing with time, as is in the radiation/matter-dominated
universes. In order to find an accelerating universe which
satisfies
\begin{equation}
\ddot{a}>0 \leftrightarrow
\frac{d}{dt}\Big(\frac{H^{-1}}{a}\Big)<0 \leftrightarrow
\omega<-\frac{1}{3},
\end{equation}
we need a shrinking Hubble scale, as was shown in the inflationary
universe. It means  that the changing rate of $H^{-1}/a$ with
respect to $a$ is always negative for an accelerating universe.
For this purpose, we introduce the future event horizon $L_{\rm
\Lambda}=R_{\rm h}=a \int_t^{\infty} (dt/a)=a
\int_a^{\infty}(da/Ha^2)$ for an observer~\cite{LI,FEH}. Using an
integral form of Friedman equation of $HR_{\rm h}=c$, one finds a
promising differential equation
\begin{equation}
\label{DEE}
c\frac{d}{da}\Big(\frac{H^{-1}}{a}\Big)=-\frac{1}{Ha^2}.
\end{equation}
This leads to  $\rho_{\rm \Lambda}\sim a^{-2(1-1/c)}$ with
$\omega_{\rm h}= -1/3(1+2/c)<-1/3$ which is close enough to $-1$
to agree with the data.  For $c=1$, we recover a case of
cosmological constant with $\omega_{\rm h}=1$ exactly.  We note
that the Friedman equation with the holographic energy density
$\rho_{\rm \Lambda}$ takes the form $ H= c/L_{\rm \Lambda}$,
whereas the Friedman equation with the conventional form
$\tilde{\rho}_{\rm \Lambda}$ is given by $H =\sqrt{8
\pi/3M_p^2}/\tilde{L}_{\rm \Lambda}^2$. Hence the above result
using the holographic energy density is no longer valid for the de
Sitter thermal energy density.

At this stage we mention that $L_{\rm \Lambda}=R_{\rm h}$ seems to
be rather ad hoc chosen and one thus requires establishing a close
connection between the holographic energy density and dark energy.
Actually the important fact to remark  is that  the holographic
energy density $\rho_{\rm \Lambda}=3c^2M^2_p/8\pi L_{\rm
\Lambda}^2$ is originally derived for a decelerating phase due to
radiation/matter-dominated universes. However, the dark energy
usually derives an accelerating universe. There exists a
difference between decelerating universe and  accelerating
universe. The key point is the existence of the future event
horizon in the accelerating universe. Therefore it is not
guaranteed that $\rho_{\rm \Lambda}$ is applicable even for an
accelerating universe.

\section{Holographic dark energy and vacuum fluctuations}

If the cosmological constant arises due to the energy density of
the vacuum, one needs to investigate the structure of quantum
vacuum at  large cosmological scales. The renormalization group
approach shows that the energy density depends on the scale at
which  it is probed. Suppose that the universe has endowed us  the
two independent length scales, $L_p \sim 1/M_p$ and
$\tilde{L}_{\rm \Lambda} \sim 1/\tilde{\Lambda}$~\cite{Pad1,Bjor}.
Then we construct two energy scales : the Planck energy density of
$\rho_p=M^4_p=1/L_p^4$ and the de Sitter thermal energy density of
$\tilde{\rho}_{\Lambda}=\tilde{\Lambda}^4=1/\tilde{L}_{\Lambda}^4$.
Thus $L_p$ determines the highest possible energy density in the
universe, whereas $\tilde{L}_{\Lambda}^4$ determines the lowest
possible energy density. In this picture, observation requires
enormous fine tuning as $(L_p/\tilde{L}_{\Lambda})^2 \le
10^{-123}$. As the energy density of normal matter/radiation drop
below $L_{\Lambda}$, the thermal ambience of the de Sitter phase
will remain constant and provide the vacuum noise. Then the dark
energy  may be  the given by the geometric mean of two scales in
the universe: $\rho_{\rm DE}=\sqrt{\rho_p
\tilde{\rho}_{\Lambda}}=M_p^2/\tilde{L}_{\Lambda}^2$ which looks
like the holographic energy density $\rho_{\rm
\Lambda}=3c^2M^2_p/8\pi L_{\rm \Lambda}^2$. On the other hand, the
hierarchy of the two scales has the pattern
\begin{equation}
\label{PAT} \rho_{\rm vac}= \frac{1}{L_p^4} +
\frac{1}{L_p^4}\Big(\frac{L_p}{\tilde{L}_{\Lambda}}\Big)^2 +
\frac{1}{L_p^4}\Big(\frac{L_p}{\tilde{L}_{\Lambda}}\Big)^4+
\cdots,
\end{equation}
where the first term is the bulk energy density that needs to be
renormalized away, the second term is due to the vacuum
fluctuations, and the third one is the de Sitter thermal energy
density.

 We will show that the holographic energy density
$\rho_{\rm \Lambda}=3c^2M^2_p/8\pi R_{\rm h}^2$ could be generated
by the vacuum fluctuations of the energy density. If the
accelerating universe has the future event horizon (the
cosmological horizon) that blocks information, the natural scale
is given by the size of the horizon $R_{\rm h}$. The operator
${\cal H}(< R_{\rm h})$, corresponding to the total energy inside
a region bounded by a cosmological horizon, will exhibit
fluctuations $\Delta E$, because the vacuum state is not an
eigenstate of ${\cal H}(< R_{\rm h})$. The corresponding
fluctuation in terms of the energy density is given by
\begin{equation}
\Delta \rho \sim \frac{\Delta E}{R_{\rm h}^3} \equiv f(L_p,R_{\rm
h}),
\end{equation}
where $f$ is a function to be determined.  In order that $\Delta
\rho \sim M_p^2/R_{\rm h}^{2}$, it is necessary to have $ (\Delta
E)^2 \sim R_{\rm h}^2/L_p^4$. This means that the square of energy
fluctuations should scale as the enclosed surface of the
accelerating universe. Actually a calculation~\cite{Pad2} showed
that for $R_{\rm h} \gg L_p$, $ (\Delta E)^2 =C_1 R_{\rm
h}^2/L_p^4$ where $C_1$ depends on the UV cutoff. Hence we roughly
prove that $\rho_{\rm \Lambda} \sim \Delta \rho$. This means that
the holographic energy density deriving an accelerating universe
with the future event horizon could  be  given by the vacuum
fluctuations of the energy density.

\section{Conclusion}
We show that the holographic energy bound $\rho_{\rm
\Lambda}=3c^2M^2_p/8\pi R_{\rm h}^2$ proposed by Cohen {\it et al}
can be  derived from the cosmic holographic bound, the
Bekenstein-Hawking bound for a weakly gravitating system. If the
IR cutoff is chosen by the future event horizon, then  the
holographic energy density can derive  an accelerating universe.
In this case  the holographic energy density   could be given by
vacuum fluctuations of the energy density.

\section*{Acknowledgment}
This work  was supported in part by KOSEF, Project Number: R02-2002-000-00028-0.

\end{document}